# "Connected Researches" in "Smart Lab Bubble": A lifeline for Commercial Agriculture in "new normal"

Prabath Chaminda Abeysiriwardana[1*], Udith K. Jayasinghe-Mudalige[2], Saluka R. Kodituwakku[3]

1*. Corresponding Author, Post Graduate Student, Department of Agribusiness Management, Faculty of Agriculture and Plantation Management, Wayamba University of Sri Lanka, Makandura, Gonawila (NWP) and Assistant Director of State Ministry of Skill Development Vocational Education Research and Innovation, Sethsiripaya Stage I, Battaramulla, Sri Lanka, abeysiriwardana@yahoo.com ; pcabeysiriwardena@wyb.ac.lk ORCID iD: 0000-0003-0300-2020

2. Professor, Vice Chancellor, Wayamba University of Sri Lanka, Makandura, Gonawila (NWP)

udith@hotmail.com ; udith@wyb.ac.lk

3. Professor, Department of Statistics & Computer Science, Faculty of Science, University of Peradeniya, Peradeniya (20400), Sri Lanka

salukak@pdn.ac.lk ; salukak@sci.pdn.ac.lk

**Abstract**

**Purpose:** Research in commercial agriculture is the best strategy that can be adopted by a country to keep on track of second sustainable goal – zero hunger by 2030. Analysing the drawbacks of present research environment and find solutions through digital intervention would be ideal solution to de-isolate the research out come in light of disruptions caused by the Covid pandemic.

**Findings:** Performance of the research institutes is not expected to remain the same and would prefer to be stagnated at a lower level. The right evacuation plan that could be work out by establishing a connected research through digital solution and followed by digitally endorsed performance monitoring and evaluation would be saviour for keeping the research in commercial agriculture live at this pandemic.

**Design/methodology/approach:** This paper will discuss what are the problems in carrying out research in commercial agriculture and propose a conceptual model to connect research beyond physical presence by digital transformations in organization design of research institutes in light of Covid-19. Further, digitally endorsed performance measurements and evaluation is envisaged in a digitally empowered connected lab complex - "Smart Lab Bubble" that is further facilitated through policy measures.

**Originality:** The connected lab complex called the "Smart Lab Bubble" concept we present here could be viewed or applied in different perspectives to engineer the real need of the time for the sustainability of research in commercial agriculture. Further, it could be adopted in research in other life science areas.

1. Importance of Research Institutes in Commercial Agriculture

The contribution from commercial agriculture could play major role in achieving the no. 2 Sustainable Development Goal (SDG) - Zero Hunger by 2030 (Sunderland, et al., 2019) (Guenette, 2019) (Gassner, et al., 2019) (UNCTAD, 2017) (Bhavani & Rampal, 2020). Targets associate with agriculture initiatives to reduce poverty and hunger and enhancing economic and household income growth are set by SDGs 2, 3, 6, 7, 8, 12, 13,14,15, 17 (FAO, 2019)

In spite of this, records show world is off track in ensuring access to safe, nutritious and sufficient food for all people all year round, and eradicating all forms of malnutrition (FAO; IFAD; UNICEF; WFP; WHO, 2020). The task of research institutes work for agriculture becomes even more challenging with an additional threat (Yasmi, Dawe, Zhang, Balie, & Dixie, 2020) (Department of Economic and Social Affairs, 2020) isolation of research institutes amid COVID pandemic.

2. Challenges of Research Institutes confronted by COVID 19

Researches are carried out by academicians in universities, researchers in research institutes, employees in labs of research unit of commercial product/ service development organizations and students or scholars either in above any institutes.

There are three kind of researches that could be observed in this situation

1. Researchers who work in normal research works directly aligned with objectives of research institutes
2. Researchers who work directly on covid 19 virus and its effects (This may be in addition to their normal research agenda)
3. Researchers who have adapted their researches to coexist with study on covid 19 virus and its effects.

The survey carried out by (Walker, Brewster, Fontinha, & Haak-Saheem, 2020), it was found that researchers felt it has undermined their confidence in applying for certain grants of the other major challenges humanity faces than Covid 19. This has been more valid for fields like commercial agriculture research.

Covid-19 lockdown force researchers to adopt remote working for their research with almost no alternatives and with poor facilities. Traditional online communication increase the amount of time taken to prepare material and require more material to achieve similar tasks. Some kinds of research are not possible to conduct any other places due to limitations in the available facilities at those remotely available labs. Almost every researcher face many problems in light of the procedures to be adopted in adherence to the safety processes relevant to Covid-19 (Gewin, 2020) (Makoni, 2020) to avoid human touch.

3. **Importance of Performance Measurements**

It is simply, amount and quality of the Agriculture products and services that research should focus on to improve. In new normal, these amount and quality should be achieved with less time and low or scarce inputs aggravated by covid pandemic. All the stakeholders of the economy should be aware of the current performance of agriculture in the pandemic locked down and any trend that may need to be reversed (UNESCAP-SSWA, 2020) (OECD, 2020) (Kurth, Walker, & Subei, 2020). Therefore, more focus should be concentrated on performance of research institutes and researchers who work towards success of commercial. There is no way that we can ignore or regress the performance measurement in any situation even for a temporary period (Knight, 2020).

Challenge is the keeping performance high in research work at research institutes in all of those circumstances when legacy metrics may be misleading and unhelpful for assessing remote performance (Schrage, 2020). Hence, monitoring agricultural performance, using performance indicators, needs to receive urgent and increasing attention in order to tackle the funding, time and location constraints which are imposed on research during the pandemic (Walker,

Brewster, Fontinha, & Haak-Saheem, 2020). Question is how we should measure performance in such situation and what policy changes required enabling such performance monitoring and evaluation.

**4. Spin with digital revolution to gain momentum in remote work**

It is very well known that thousands of labs in a variety of research fields are reconsidering their planned studies and not all projects can be easily put on freeze (Servick, Cho, Guglielmi, Vogel, & Couzin-Frankel, 2020). Transfer of a research from one place to another place where the conditions are good for that particular research facilitated by cooperation and empowered by national level policy changes would be good option in these situations. This will ensure researchers ability to access alternative similar resources, which has been significantly affected by the lockdown.

It is important to introduce remote work practices suitable for local conditions after understanding the differences in the impact of the pandemic lockdown on researcher to make the impact of the lockdown on research activity is not negative for any researcher but make them feel it has opened more time to spend on their work. It should be very careful to keep human resource policies in place where balance of the assessment and the safe side of new skills and capabilities is ensured through strategies like upskilling workers amid increasing automation, while causing change of attitude of the researcher towards need of the country and world instead of gaining on academic credit.

Thriving on present situation we can suggest interlink research institutes and its lab spaces that would play a significant role in connecting and empowering researchers to do their research with effectively and efficiently and it would enable regulatory and policy environment that is required for the mass adoption of research networks. A policy change should be there to induce more resource sharing to be used in research work rather than researchers opts for dedicated equipment for their research work.

In light of this, we would propose a digitally empowered experimenting environment for researchers to find solutions to above issues and detail what measures could be adopted and how they should be empowered by digital transformations and what policy measures would enable such an environment.

**5. "Smart Lab Bubble" – connected laboratory complex for researchers of life sciences**

The total Internet of Things (IoT) connected devices is projected to the amount around 25 to 75 billion units worldwide according to several claims (Holst, 2021) (Horwitz, 2019) and each person will own 15 connected devices by 2030 (Heslop, 2019). In this context and never ending pandemic situation, we propose integrated framework for implementing smart lab services called "Smart Lab Bubble". Here, labs are supposed to interconnect real-time and around the clock through communication networks to share equipment, data & information and human resources. Here already existing smart city concept would be adopted in interconnecting existing labs into a Smart Lab Bubble empowered with IoT capabilities. Further, it would be automated with robotic and virtual experimentation when lab is out of human presence. Each Smart Lab Bubble could be developed in line with relevant stream of subject areas. For example, Chemistry, Physics, Botany, and Biotechnology or using existing lab and equipment space in the country as appropriate. Further, these bubbles may be interconnected each other to form one lab space or lab complex (Smart Big Bubble). Satellite link could be used as appropriate and as per necessitates in these bubbles for increased overall IoT network performance (Petrovic, et al., 2020).

## 6. How Smart Lab Bubble will work with IoT

An IoT enabled Smart Lab Bubble will use all devices including automated devices, such as collaborative robots or pipetting robots connected to the cloud or local server that can be controlled by the researcher externally and accessed from anywhere with an internet connection (Ryding, 2020). An administrative control panel with additional sensors in a central location could be used through the Internet to facilitate the researcher to conduct their experiments in a more productive integrated manner, without unnecessary obstructions or error-messages. If equipment fails, he can search for another similar equipment in real time, conduct that part of his research at another space, and transfer back with results to original place as required.

These bubbles permits scientists not essential to be physically present in their own lab to do their research and at the same time uses any other lab in the bubble or lab in any other bubble in the big bubble.

In addition, such integrated system will overcome barriers in the form of compliance issues or simply a lack of communication between devices and researchers and smart energy management.

## 7. How Smart Lab Bubble would help remotely isolated researchers

Automation of data collection will eliminate data loss that could happen when a scientist is too selective on what to record in his limited presence in a lab in new normal.

Virtual experimentation simulations are adequate for some experimental designs and trials to be prepared before actual experiment itself. This will reduce the time that scientist need to be physically presence in actual experiment environment as he is well prepared beforehand the actual experiment (Destino & Petrovic, 2021). Also this could be used to train and teach experimental group remotely with same level of hand on experience (Corter, et al., 2007) to do the experiment with reduced resources and wastes when involved in actual experiment. Some parts or whole experiment may be handover to the automation setup of the lab to reduce even more time required in physical presence.

IoT- enabling system solves connectivity issues between equipment better than any other system to maximise research potential of a lab and the researcher. The IoT enabled devices will connect all different elements of the laboratory from automated micro pipette to PCR machines to maintain smooth workflow of the experiment. Then, machine output can be directly streamed into a digital format, saving the scientist valuable time and effort along with eradicating the risk of human error (Chubb, 2020) (ELI LILLY AND COMPANY, STRATEOS, 2020). Therefore, experiments are executed easier, quicker and greater precision with a high degree of reproducibility, data is documented more accurately and research is more accessible to different parties.

Sensors could be used to monitor vital stats in plants and animals, recommend nutrients plans for growth of plants and animals, order prescriptions and call researchers attention if needed in status of malnutrition and spreading of diseases. Within plant or animal houses, IoT will be used to enable robots to perform remedial actions, while a virtual researcher will perform administrative services. With added machine learning and AI technology, these actions could be streamlined and improved for best performance. In this way, it connects every tool in the laboratory, creating a truly smart, productive environment where machines can predict experiment outcomes and support more contextual decision making using real-time data for researches.

8. **Interventions and Policies need to secure digital experimenting environment**

We propose following form of interventions and characteristics supported by frontier technologies to incorporate to the "Smart Lab Bubble".

**Real time interactive online platform for promoting research findings**

This would facilitate more transactions and trades to culture remote practice in research work in both researcher and his consumer.

**Study the possibility of using virtual reality and augmented reality in remote collaboration**

This would make research continuous through providing inputs by principle researchers (Yu, et al., 2010) to researches, which are supposed to be facilitated by local unspecialized third parties in isolated areas where main researches are locked down due to pandemic situation.

**Use of automated systems such as intelligent environment control systems remotely handled via IoT and Drone technologies**

Best-of-breed IoT platform seamlessly connects a wide-range of assets and delivers real-time performance data which supports real time decision making to prevent maintenance issues before they become maintenance problems (PTC, 2020). Ex: These technologies integrated more with precision agriculture would give decision supportive data for performance management of research in commercial agriculture.

**Promote digital innovation through policy changes such as developing more apps to thrive on existing web and internet facilities**

Use of these apps to support commercial agriculture research may act as a game-changer for the performance management of such institutes. Ex: Work Management Software integrated with performance management tools in Remote Workforces Environment.

**Collaborative scenario planning supported by digital capabilities**

Alternatively, Collaborative scenario planning done through digital capabilities enhanced by data from new types of sensors, e-marketing platforms, food traceability systems, remote sensing, earth observation services and social media, and the means to analyse this data using advanced analytic tools and artificial intelligence techniques, would improve an organization's core processes for creating supply chain uninterruptedly at a rapid pace (Joglekar & Phadnis, 2020).

**Key Performance Indicators (KPIs) innovatively geared by digital accountability**

When innovation drives the engine of research, Key Performance Indicators (KPIs) innovatively geared by digital accountability and giving due acknowledgement and respect to distinctions between work and home life will play major role in performance measurement of research institutes works towards commercial agriculture in light of this pandemic period and may be good option in future as well. As researchers are dispersed and distanced, using KPIs associated with individuals rather than with team, play effective and efficient role in performance management and result in more transparency for research performance. Research institutes may introduce and renovate if not exist their data-driven dashboards to better inspire people and project teams and promote positive outcomes. They must automatically capture and analyse, and explicitly communicate, their high-performance criteria to generate real-time analytic insights (Schrage, 2020). Personalized KPIs integrate with an app and give real time pulse of present situation of the performance will act as a blood pressure meter of a researcher in a particular institute and thus very friendly make researchers works towards common goals of success in commercial agriculture for zero hunger in 2030 even in pandemic era.

To exercise an extreme form of social distancing while at work and make it not complicated for group work; new form of work schedules should be prepared and guidelines must be developed on how research activities should be carried out safely with the aid of real time data mining supported by frontier technology interventions.

## 9. Conclusion

We believe that the process used innovate inherently fuelled by digital pulses would be best suited to discover, create and develop ideas, to refine them into useful forms and to use them to earn profits, increase efficiency and/or reduce costs in commercial agriculture research in a locked down situation (Morris, 2020) along with the proposed "Smart Lab Bubble" experimenting environment. Leaders of research institutes in commercial agriculture must carefully scrutinize landscape of emerging technologies and strategically leverage innovation to transform how their research involve with long term sustainable economic and social impacts. In other words, they must put more weight on process-focused innovation instead of product-focused innovation or, as it is better known, business model innovation (BMI) to trade off risk and return of research in this pandemic situation (Hasija, 2020).

This is an exciting time full of unanticipated and disconcerting lines of developments, but everyone know that it is not always easy to change and match everything in new normal. It is high time leveraging digital technology to create new research values out of performance measurements. Therefore, strategy changes and policy interventions in innovative way in the form of digital transformation are essential for research leadership to close the gap between research organizations that are digitally mature and those struggling to adapt it for success of commercial agriculture in the face of seemingly never ending locked down.

**Declarations**

**Acknowledgments**

I thank in advance the anonymous reviewers for their helpful comments.

**Financial and Non financial interests**

The authors have no relevant financial or non-financial interests to disclose.

**Conflict of interest**

We know of no conflicts of interest associated with this publication, and there has been no outside funding for this work that could have influenced its outcome.

**Author contribution**

We would like to present the Authors' contributions as follows: author 1 contribute to the conception/ design and draft the work and author 2 and author 3 contribute by revising it. As Corresponding Author, I confirm that the manuscript has been read and approved for submission by all the named authors.